\documentclass[aps,prl,superscriptaddress,showpacs,showkeys,a4paper,10pt,notitlepage,twocolumn]{revtex4-1}

\usepackage[utf8]{inputenc}
\usepackage[english]{babel}
\usepackage{amsmath}
\usepackage{amssymb}
\usepackage{amsfonts}
\usepackage{graphicx}
\usepackage{cancel}
\usepackage{txfonts}
\usepackage{upgreek}
\usepackage{mathtools}
\usepackage[T1]{fontenc}

\usepackage{xcolor}
\definecolor{link_blue}{RGB}{52,46,157}

\usepackage{longtable}
\usepackage[colorlinks,citecolor=link_blue,linkcolor=link_blue,urlcolor=link_blue]{hyperref}
\usepackage[tiny,center,uppercase]{titlesec}

\setcitestyle{super}

\renewcommand{\vec}{\boldsymbol}

\newcommand{\thp}[2]{\vec #1\cdot\vec #2}

\newcommand\ri{\mathrm{i}}

\DeclareMathOperator{\re}{Re}
\DeclareMathOperator{\im}{Im}

\begin{document}
	
\title{Parametric M\"{o}ssbauer radiation source}

\author{O.\ D.\ Skoromnik}
% \email[Corresponding author: ]{olegskor@gmail.com}
\email[]{olegskor@gmail.com}
\affiliation{Max Planck Institute for
  Nuclear Physics, Saupfercheckweg 1, 69117 Heidelberg, Germany}

\author{I.\ D.\ Feranchuk}
\email[Corresponding author: ]{ilya.feranchuk@tdtu.edu.vn}

\affiliation{Atomic Molecular and Optical Physics Research Group,
  Advanced Institute of Materials Science, Ton Duc Thang University, 19
  Nguyen Huu Tho Str., Tan Phong Ward, District 7, Ho Chi Minh City,
  Vietnam}

\affiliation{Faculty of Applied Sciences, Ton Duc Thang
  University, 19 Nguyen Huu Tho Str., Tan Phong Ward, District 7, Ho
  Chi Minh City, Vietnam}

\affiliation{Belarusian State University, 4 Nezavisimosty Ave.,
  220030, Minsk, Belarus}

\author{J.\ Evers}
\affiliation{Max Planck Institute for
  Nuclear Physics, Saupfercheckweg 1, 69117 Heidelberg, Germany}

\author{C.\ H.\ Keitel}
\affiliation{Max Planck Institute for
  Nuclear Physics, Saupfercheckweg 1, 69117 Heidelberg, Germany}

\begin{abstract}
  Numerous applications of M\"{o}ssbauer
  spectroscopy~\cite{rohlsberger_nuclear_2005, kalvius_rudolf_2012,
    jaeschke_nuclear_2016} are related to a unique resolution of
  absorption spectra of resonant radiation in crystals, when the
  nucleus absorbs a photon without a recoil. However, the narrow
  nuclear linewidth renders efficient driving of the nuclei
  challenging, restricting precision
  spectroscopy~\cite{adams_scientific_2019, heeg_spectral_2017},
  nuclear inelastic scattering~\cite{baron_introduction_2015} and
  nuclear quantum optics~\cite{adams_x-ray_2013,
    jaeschke_quantum_2016, kuznetsova_quantum_2017}. Moreover, the
  need for dedicated X-ray optics~\cite{alexeev_nuclear_2019,
    kalvius_rudolf_2012} restricts access to only few isotopes,
  impeding precision spectroscopy of a wider class of systems.  Here,
  we put forward a novel M\"{o}ssbauer source, which offers a high
  resonant photon flux for a large variety of M\"ossbauer isotopes,
  based on relativistic electrons moving through a crystal and
  emitting parametric M\"ossbauer radiation essentially unattenuated
  by electronic absorption. As a result, a collimated beam of resonant
  photons is formed, without the need for additional
  monochromatization. We envision the extension of high-precision
  M\"{o}ssbauer spectroscopy to a wide range of isotopes at
  accelerator facilities using dumped electron beams.
\end{abstract}



\maketitle

Traditional M\"ossbauer spectroscopy uses radioactive sources, which
provide essentially background-free near-resonant $\gamma$-radiation
with a spectral width of order of the natural linewidth of the
involved nuclear transitions~\cite{rohlsberger_nuclear_2005,
  kalvius_rudolf_2012, jaeschke_nuclear_2016}. Accelerator-based X-ray
sources offer orders of magnitude more resonant photon flux, but the
short X-ray pulses contain an intense off-resonant background, which
strongly exceeds the resonant component. As a result, M\"ossbauer
spectroscopy usually is performed in the time
domain~\cite{jaeschke_nuclear_2016, ruby_mossbauer_1974}, removing the
``prompt'' non-resonant background via temporal gating of the
detectors. This, for instance, restricts the study of short-lived
isotopes, for which the gating leads to a severe loss of signal
photons due to the fast initial decay.

Alternatively, synchrotron M\"ossbauer sources
(SMS)~\cite{gerdau_nuclear_1985, PhysRevB.55.5811,
  masuda_development_2008, PhysRevLett.102.217602, potapkin_57fe_2012}
can be employed to monochromatize the synchrotron radiation to few
natural linewidths using pure nuclear Bragg reflexes, enabled by the
suppression of electronic reflections via particular crystal
symmetries. In addition, usually a specific M\"ossbauer isotope is
targeted, requiring dedicated X-ray optics such as monochromators to
reduce the off-resonant background component. Therefore, it is
challenging to make new M\"ossbauer isotopes accessible at modern
pulsed X-ray sources, which hinders the exploration of new scientific
applications of specific M\"ossbauer
nuclei~\cite{alexeev_nuclear_2019, kalvius_rudolf_2012}.

\begin{figure*}[t]
  \includegraphics[width=0.8\textwidth]{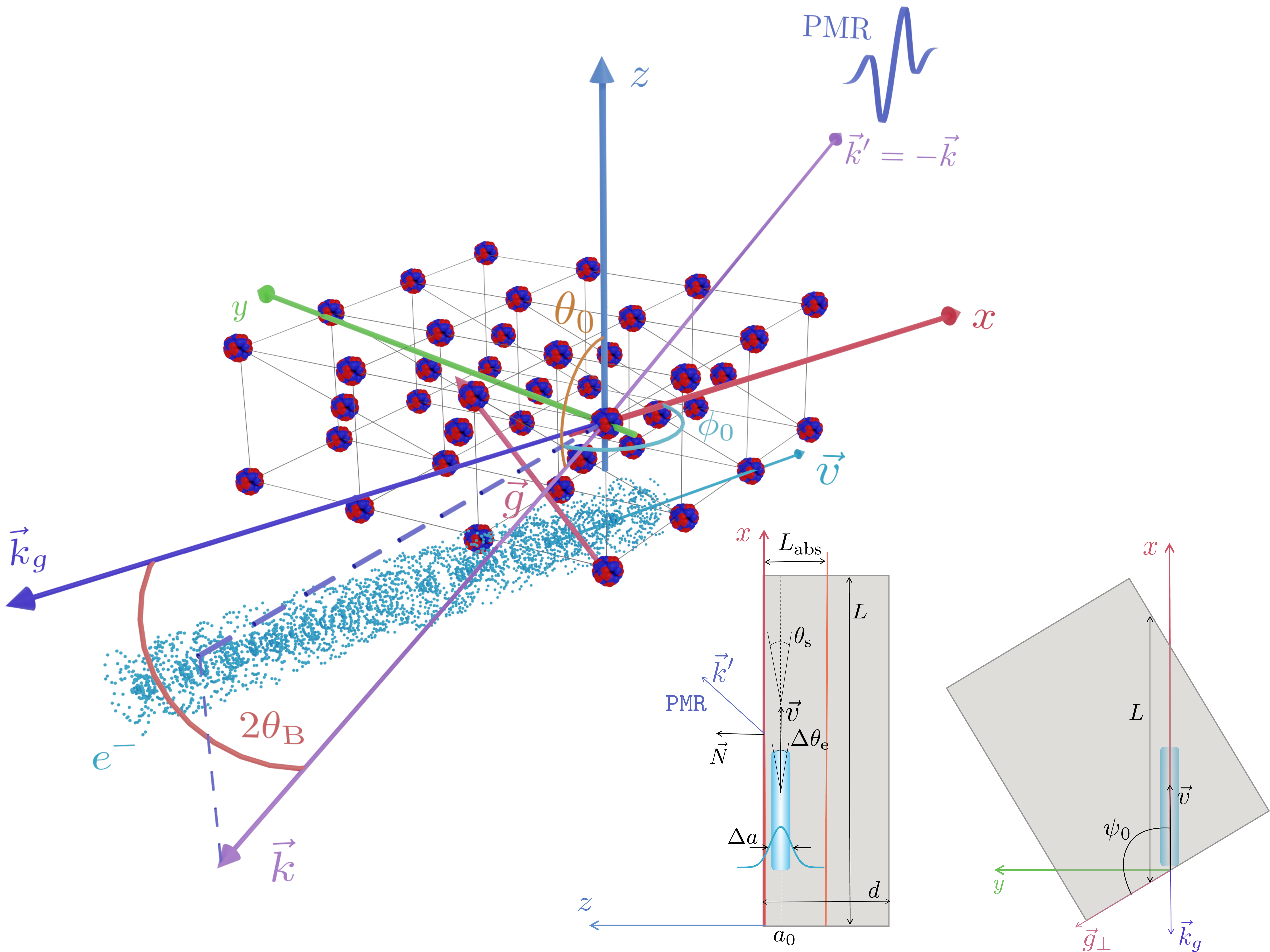}
  \caption{{\bf Schematic setup of the PMR generation.} The electron
    bunch moves uniformly with velocity $\vec{v}$ in the $x$
    direction. The crystal surface lies in the $x-y$ plane.  $\vec{g}$
    is the reciprocal crystal lattice vector. PMR will be mainly
    emitted in the direction given by the vector
    $\vec k' = -\vec k = -k_{0}(\sin\theta_{0} \cos\phi_{0},
    \sin\theta_{0} \sin\phi_{0}, \cos\theta_{0})$, and thus rapidly
    leaves the crystal without significant electronic
    absorption. $\vec k_{g} = \vec k + \vec g$ and
    $\theta_{\mathrm{B}}$ is the Bragg angle. The energy of the PXR is
    tuned using the angle $\psi_0$ between the electron velocity and
    the projection $\vec g_{\bot}$ of $\vec{g}$ on the crystal
    surface. The resonance condition is
    $\vec k^{2} = (\omega_{0}/c)^{2} = k_{0}^{2}$, which yields
    $\cos\psi_{0} = -v/(k_{0} c) \times (k_{0}^{2}\gamma^{-2} + g^{2})
    / (2\sqrt{g^{2} - g_{z}^{2}})$.}\label{fig:1}
\end{figure*}

An alternative scheme to generate X-rays is parametric X-ray radiation
(PXR), based on relativistic electrons moving through a
crystal~\cite{PXR_Book_Feranchuk,BARYSHEVSKY1986306,
  J.Phys.France1985.46.1981, Brenzinger1997, PhysRevLett.79.2462,
  PhysRevAccelBeams.21.014701}. In PXR, the electron self-field
diffracts on the crystallographic planes, which leads to the
generation of electromagnetic radiation. Its relative spectral and
angular widths are suppressed by the large electron energy $E$, via
the relativistic $\gamma$ factor $\gamma = E/m_{\mathrm{e}}c^{2}$,
resulting in quasi-monochromatic and well collimated PXR
radiation. Moreover, it is possible to fix the electron angle of
incidence in such a way that one of the PXR peaks is in resonance with
a nuclear M\"ossbauer transition, giving rise to Parametric
M\"ossbauer Radiation (PMR)~\cite{ahmadi_parametric_2013}.

However, conventional PXR schemes are limited in intensity due to
substantial X-ray absorption in the
crystal~\cite{ahmadi_parametric_2013,AHMADI201378}. This can be
understood by noting that the PXR intensity depends on the crystal
polarizability~\cite{J.Phys.France1985.46.1981,PXR_Book_Feranchuk}. For
crystal diffraction, the polarizability is maximized near the
resonance frequencies, where also the absorption becomes large. To
overcome this issue, a particular geometry featuring extremely
asymmetric diffraction (EAD) was
suggested~\cite{SKOROMNIK201786}. This geometry exploits a peculiar
PXR feature, namely, that the radiation is emitted under a large angle
relative to the electron velocity, which is in stark contrast with
other mechanisms generating radiation from relativistic particles. In
the EAD geometry, the electrons are moving in a thin crystal layer
parallel to the crystal-vacuum interface in such a way that the
emitted photons immediately exit the crystal without much
absorption. This effectively increases the intensity of the radiation
by two orders of magnitude with respect to the conventional transition
geometries. However, such EAD geometries have not been studied in the
case of PMR.

Here, we put forward a novel versatile source for M\"{o}ssbauer
spectroscopy, which is based on PXR simultaneously satisfying the
M\"{o}ssbauer resonance condition to effectively excite the nuclei and
the EAD condition to suppress absorption. This source offers a
competitive nuclear resonant photon flux for a large variety of
M\"{o}ssbauer crystals. In addition, for certain crystals the SMS
crystal symmetry condition can be fulfilled leading to the suppression
of the off-resonant electronic background radiation. In this case, our
calculations predict almost background-free emission of M\"ossbauer
radiation, paving the way to M\"{o}ssbauer spectroscopy on short-lived
isotopes directly in the energy domain, without the need for
additional time gating or the development of dedicated
monochromatizers.

We illustrate our approach in the case of ${}^{121}$Sb, for which our
simulations predict about ~$10^3$ resonant photons per second and
nuclear linewidth, essentially background-free. We further discuss two
isotopes without the SMS condition: ${}^{133}$Cs in order to
illustrate the interplay between the electron and nuclear components
of the crystal polarizabilities, and ${}^{57}$Fe as the classical
workhorse of M\"{o}ssbauer spectroscopy. In the latter case, more than
$10^4$ photons per second and natural linewidth $\Gamma$ are
predicted. The total linewidth of a PMR source is $\sim 5-20\ \Gamma$
depending on the crystal.
\begin{figure*}[t]
  \includegraphics[width=0.48\textwidth]{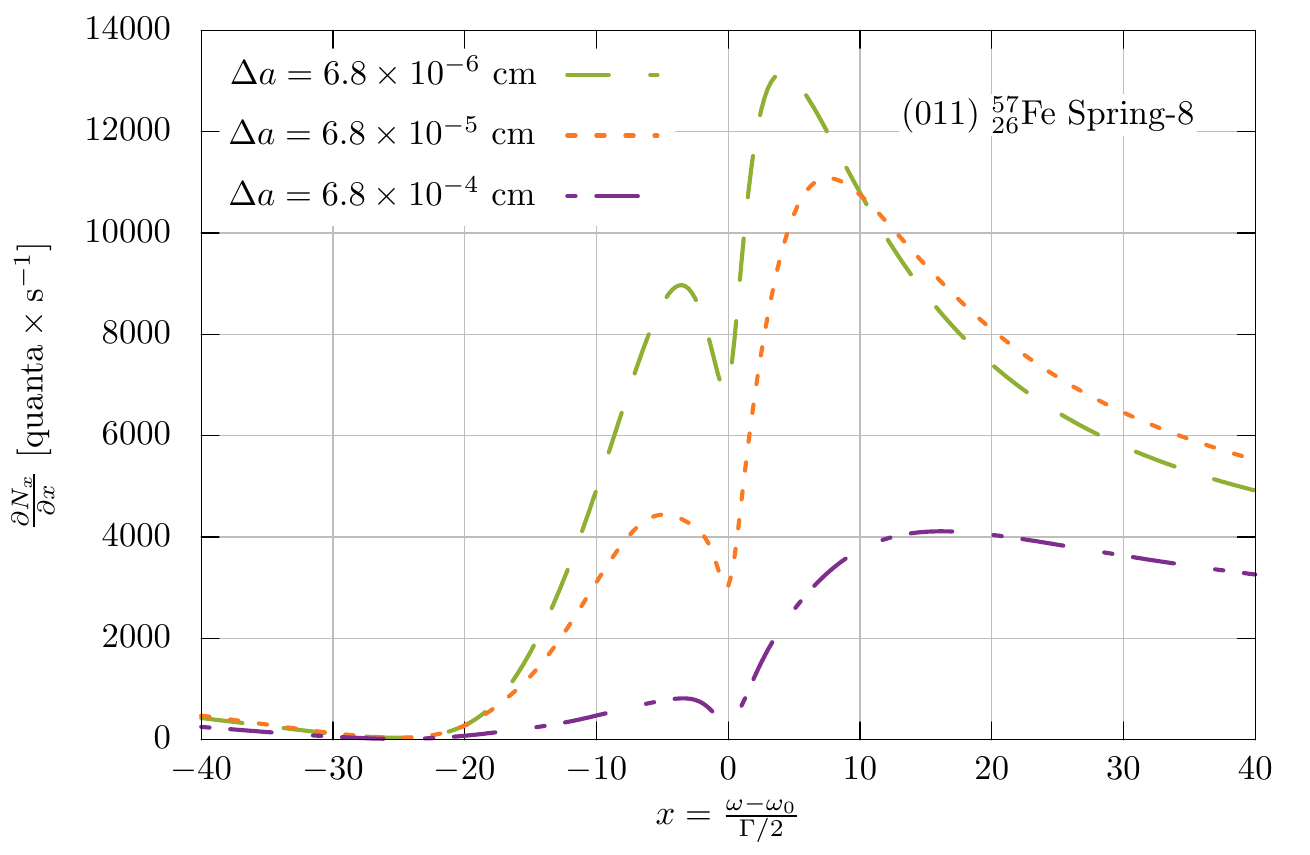}
  \includegraphics[width=0.48\textwidth]{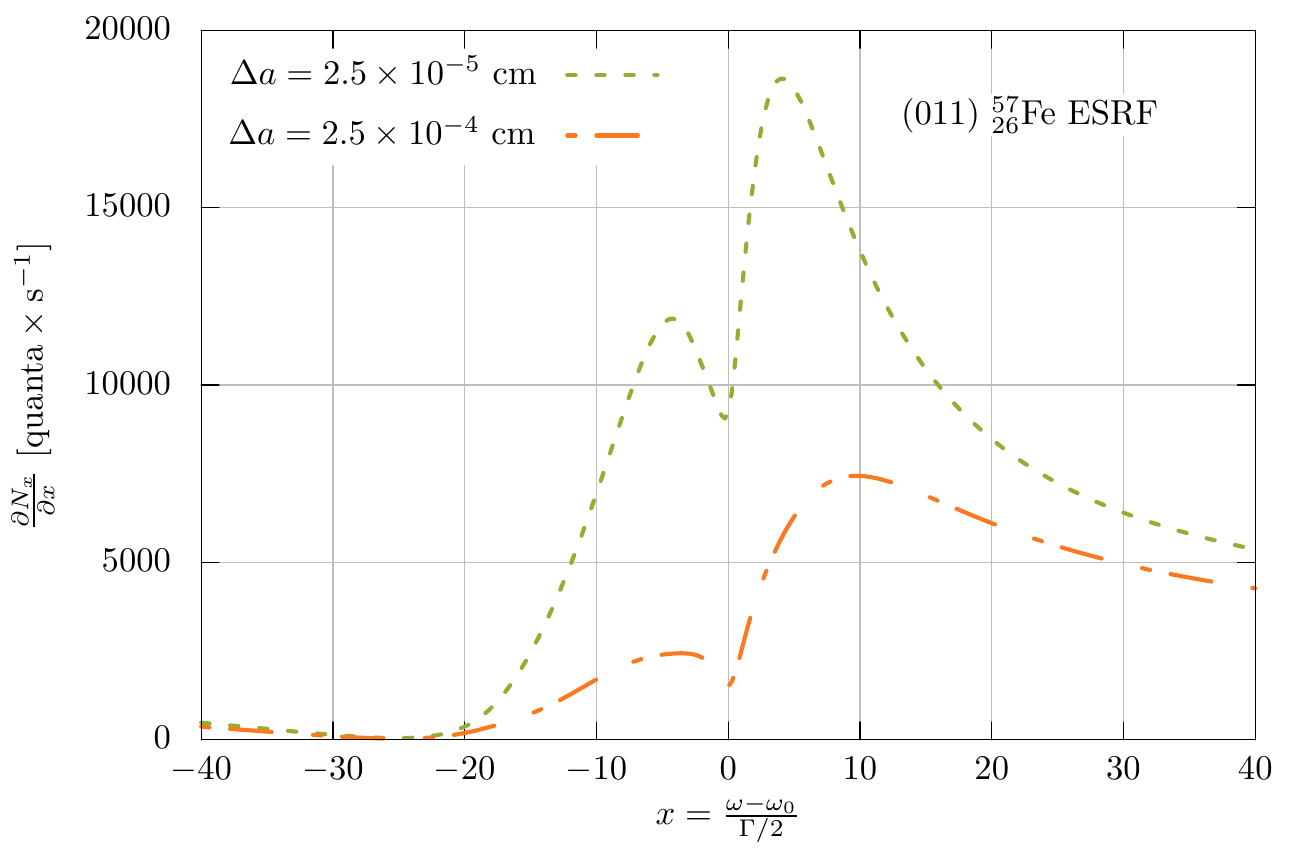}
  \\
  \includegraphics[width=0.48\textwidth]{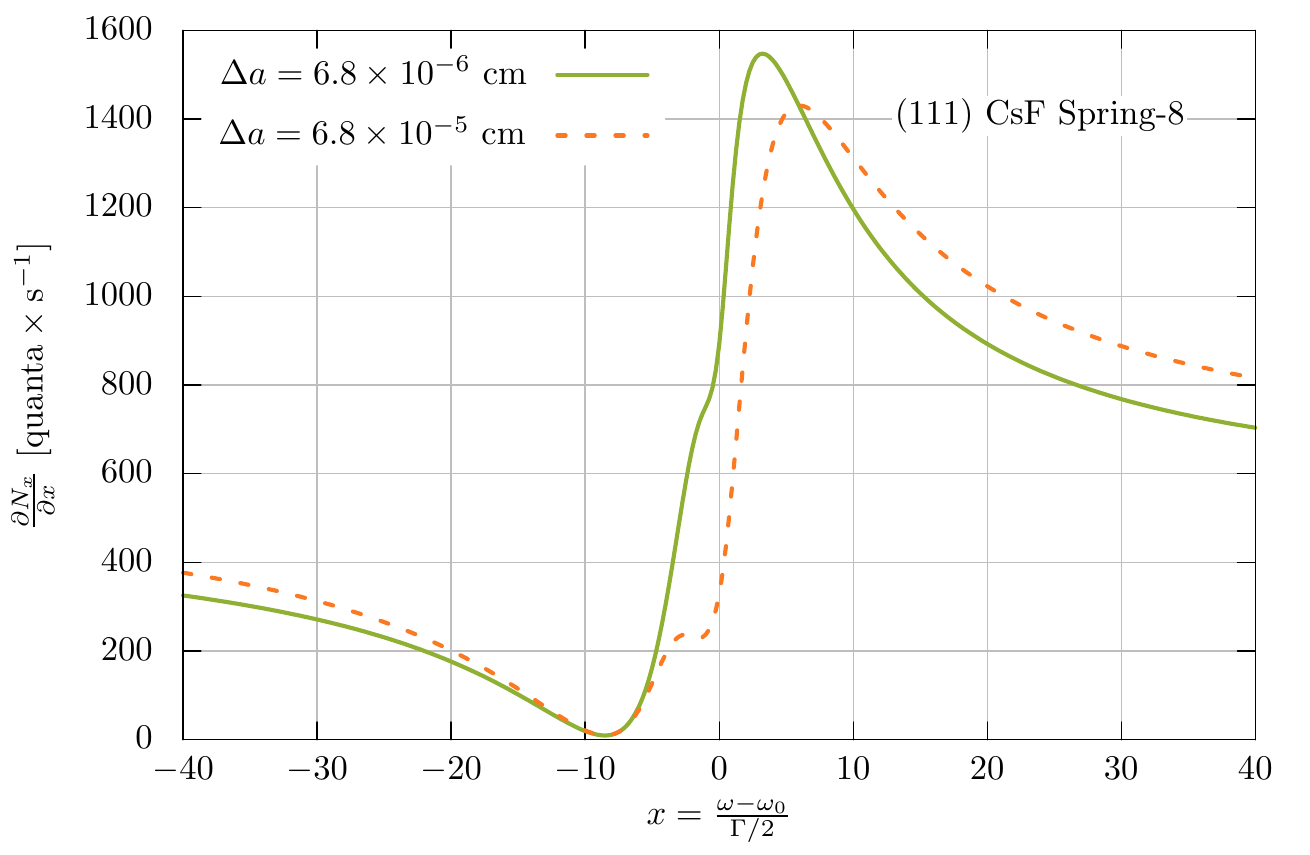}
  \includegraphics[width=0.48\textwidth]{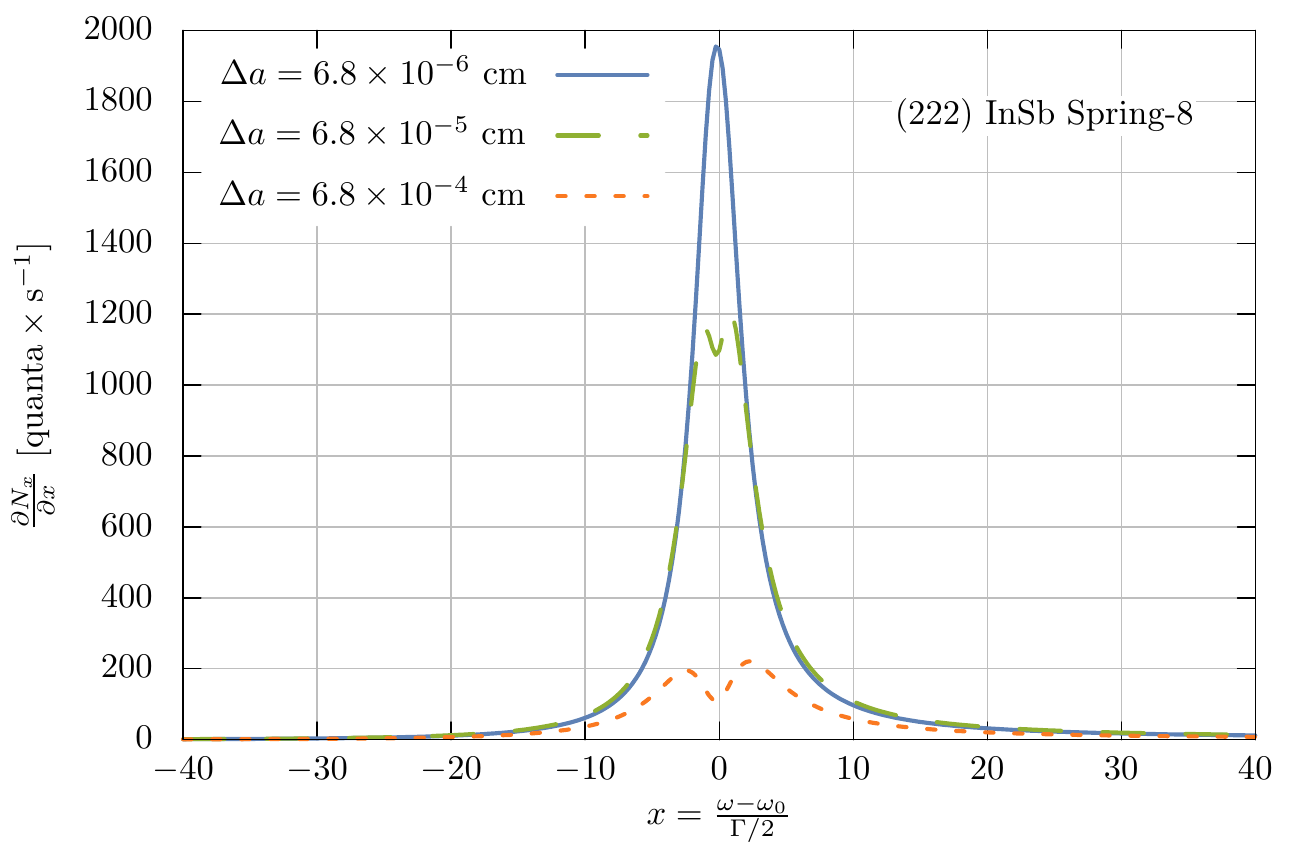}
  \caption{{\bf Emission spectra of the PMR source.} The figure shows
    the number of emitted X-ray photons per second as a function of the
    dimensionless frequency $x = (\omega - \omega_{0}) /
    (\Gamma/2)$. The results are averaged over the electron beam
    parameter distributions, for different transversal widths and
    divergences of the electron bunch, keeping the emittance constant.
    The top row compares the $(011)$ reflection of an $\alpha$-iron
    crystal for Spring-8 and ESRF electron beam parameters.  The
    bottom left panel shows the emission from the $(111)$ reflection
    for the CsF crystal, with Spring-8 electron parameters.  The
    bottom right panel shows pure PMR emission from the $(222)$
    reflection of the InSb crystal.  For all panels, we assume an
    angular spread in the $y$ direction of $10^{-3}\ \mathrm{rad}$.
    The crystal lengths are chosen as $L = 0.5\ \mathrm{cm}$ (Spring-8)
    and $L = 0.2\ \mathrm{cm}$ (ESRF),
    respectively.
    % \textbf{Parameters of the Spring-8 and ESRF
    %   facilities.}
    At Spring-8, the electron energy is $E=8000$~MeV, the vertical
    emittance is $\epsilon = 6.8\times 10^{-10}$ cm$\cdot$rad, the
    vertical beam size is $\Delta a = 6\times10^{-4}$~cm, and the
    electron current is $j = 100\ \mathrm{mA}$. At ESRF, the
    corresponding parameters are $E = 6030$~MeV,
    $\epsilon = 2.5\times 10^{-9}$~cm$\cdot$rad,
    $\Delta a = 7.9\times 10^{-4}$~cm, and $j =
    200$~mA. }\label{fig:2}
\end{figure*}

In order to calculate PXR and PMR (see Methods for details on the
calculation), we solve the inhomogeneous Maxwell's equations using a
Green's function and the standard two-wave approximation approach of
dynamical diffraction theory~\cite{authier2001dynamical}. We find that
the wave vector $\vec k' = -\vec k$ corresponding to the maximum PXR
emission is determined as a solution of two equations: (a) the
Cherenkov radiation condition~\cite{BARYSHEVSKY1986306}
\begin{align}
  q' = \re q = 1 + \frac{(\vec k + \vec g) \cdot \vec v}{\omega_{0}} =
  0 \label{eq:1}
\end{align}
for the diffracted wave and (b) minimal value for the deviation from
Wulff-Bragg's condition\cite{BARYSHEVSKY1986306,
  J.Phys.France1985.46.1981,PXR_Book_Feranchuk}
\begin{align}
  |\alpha_{\mathrm{B}}|= \frac{|(\vec k + \vec g)^{2} -
  \vec k^{2}|}{|\vec k|^{2}} = \frac{|2\thp{k}{g} + g^{2}|}{|\vec
  k|^{2}} \label{eq:2}
\end{align}
The latter condition describes the diffraction of an electron
self-field on the crystallographic planes with the reciprocal lattice
vector $\vec g = (g_{x}, g_{y}, g_{z})$, where $\vec v$ is the
electron velocity and $|\vec k| = \omega_{0}/c$ with $\omega_{0}$ the
frequency of the resonant M\"ossbauer transition.

Next, we consider the EAD geometry case~\cite{SKOROMNIK201786,
  skoromnik_parametric_2019}, see Fig.~\ref{fig:1}, in which electrons
are moving parallel to the crystal-vacuum interface (parallel to $x-y$
plane) and emit radiation under a large angle to the crystal
surface. In this geometry, the angle $\psi_{0}$ between
$\vec g_{\bot}$ and the electron velocity $\vec v$ we adjust in such a
way that the frequency of the emitted radiation is coincident with the
resonance frequency of the M\"{o}ssbauer isotope.

\begin{figure}[t]
  \includegraphics[width=0.48\textwidth]{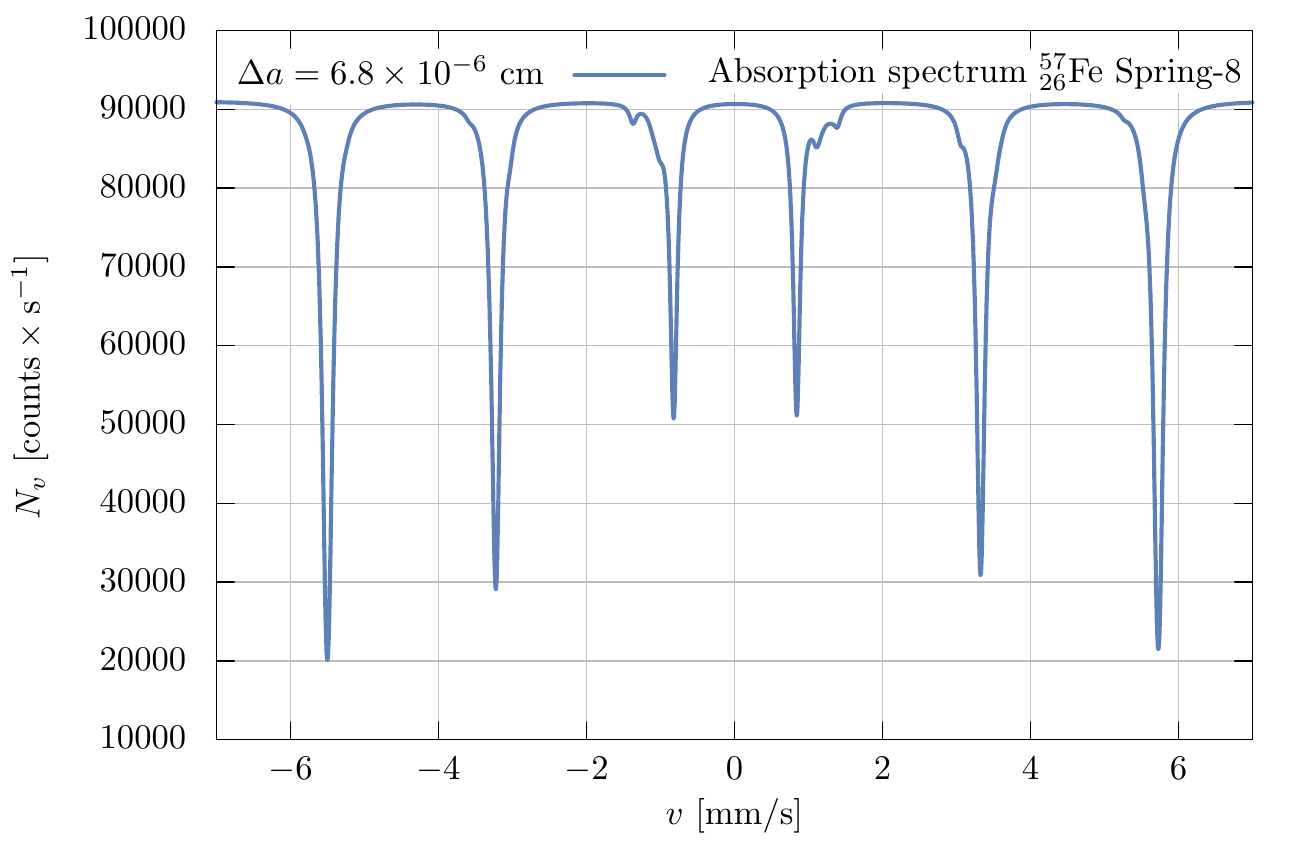}
  \caption{\label{fig:3}{\bf Example absorption spectrum.} An
    absorption spectrum simulated with the PMR predicted for the
    Spring-8 experimental facility with
    $\Delta a = 6.8 \times 10^{-6}$cm. As a target, a non-enriched
    ($\eta = 0.02$) $\alpha$-iron crystal of thickness
    $D = 5\ \mu\mathrm{m}$ is assumed. The contrast is determined via
    the ratio of the electron and the nuclear contributions into the
    crystal polarizability near the resonance frequency and in the
    case of $\alpha$-iron $\sim 20$.}
\end{figure}

Solving Eqs.~(\ref{eq:1})-(\ref{eq:2}) under the condition
$\vec k_{\vec g}\cdot\vec N = (\vec k + \vec g) \cdot \vec N = 0$,
which specifies the EAD geometry, we find that the maximum of the
X-ray emission is in the direction
$\vec k' = -\vec k = (g_{x} + \omega_{0}/v, g_{y}, g_{z})$. The $z$
component causes the generated radiation to immediately leave the
crystal, such that absorption within the crystal is greatly reduced.

Next, we insert the solution of the diffraction problem for the
electric field in the standard expression for the energy density of
the emitted radiation and integrate over the particle trajectory and
over the X-ray spherical emission angle $\phi$. This yields the
spectral-angular distribution of the emitted photons
\begin{align}
  \frac{\partial^{2} N}{\partial \omega \partial \theta}
  &= \frac{e_{0}^{2}}{4\pi \hbar c \omega}\frac{1}{|\sin\phi_{0}|}
    \sum_{s = \sigma,\pi} \Bigg(|E_{g1s}|^{2}
    \left(\frac{\thp{v}{e_{1s}}}{c}\right)^{2}  \nonumber
  \\
  &\mspace{120mu}\times \frac{1 - e^{-2 k_{0}L
    q_{s}''}}{q_{s}''} e^{-2 k_{0} |z_{0}
    \varepsilon_{1s}''|}\Bigg), \label{eq:3}
\end{align}
where $L$ is the crystal length, $z_{0}$ is the electron initial
coordinate, and
$E_{g1s} = c_{s}\chi_{\vec g} / (\alpha_{\mathrm{B}} + \chi_{0})$ is
the amplitude of the diffracted wave.
The index $s$ sums over the $\sigma$ and $\pi$ polarizations, with
$c_{\sigma} = 1$ and $c_{\pi} = \cos2\theta_{\mathrm{B}}$, and
polarization vectors
$\vec e_{1\sigma} = \vec k \times \vec g / |\vec k \times \vec g|$ and
$\vec e_{1\pi} = \vec k_{g}\times \vec e_{1\sigma} / |\vec k_{g}\times
\vec e_{1\sigma}|$.
Finally, $q_{s}'' = |\theta_{\mathrm{e}z} \varepsilon_{1s}''|$, where
$\theta_{\mathrm{e}z}$ characterizes the $z$-component of the electron
velocity, and $\varepsilon_{1s}''$ is the imaginary part of the
solution of the dispersion equation~\cite{skoromnik_parametric_2019}
for the fields in a crystal
$\varepsilon_{1s} = -\chi_{0}/(2\cos\theta_{0}) + c_{s}^{2}\chi_{\vec
  g}\chi_{-\vec g} / [2 (\alpha_{\mathrm{B}} + \chi_{0})
\cos\theta_{0}]$.

This expression contains two key quantities, which determine PMR and
PXR, namely, the dielectric susceptibilities
$\chi_{0}(\omega) = \chi_{0\mathrm{e}}(\omega_{0}) +
\chi_{0\mathrm{n}}(\omega)$ and
$\chi_{\vec g}(\omega) = \chi_{\vec g\mathrm{e}}(\omega_{0}) +
\chi_{\vec g\mathrm{n}}(\omega)$. They each comprise an electronic
($\chi_{0\mathrm{e}}$, $\chi_{\vec g\mathrm{e}}$) and a nuclear
($\chi_{0\mathrm{n}}$, $\chi_{\vec g\mathrm{n}}$) contribution, and
PMR becomes sizable, if the parameter
$\xi = |\chi_n(\omega_0) / \chi_e(\omega_0)|> 1 $. The nuclear part
\begin{align}
  \chi_{\vec g\mathrm{n}}(\omega) = - \frac{4\pi}{\omega_{0}^{2}
  c^{3}}\frac{S(\vec g)}{V} \frac{\eta e^{-W(\vec k, \vec
  k_{g})}}{\omega_{0}(1 + \alpha_{c})} \frac{\Gamma / 2}{(\omega -
  \omega_{0}) + \ri \Gamma / 2} \label{eq:4}
\end{align}
has a resonance character~\cite{afanasev_suppression_1965,
  rohlsberger_nuclear_2005} and is responsible for the PMR. Here,
$S(\vec g)$ is the structure factor, $e^{-W(\vec k, \vec k_{g})}$ the
Debye-Waller factor, $V$ the volume of the unit cell, $\alpha_{c}$ the
coefficient of the internal conversion and $\Gamma$ the natural line
width of the transition.

For our numerical analysis, we choose electron beam parameters from
the Spring-8\cite{Spring8Parameters} and ESRF~\cite{ESRFParameters}
storage ring facilities, where we consider the possibility to focus
the electron beams to smaller electron beam diameters $\Delta a$.

We investigated the emission from three crystals with cubic
lattices. The first two are without the SMS condition: the
$\alpha$-iron crystal, enriched to $90\%$ in the resonant
M\"{o}ssbauer isotope $^{57}_{26}\mathrm{Fe}$ and the CsF crystal,
which contains $_{55}^{133}\mathrm{Cs}$.  The third crystal --- the
InSb crystal contains the resonant isotope $_{51}^{121}\mathrm{Sb}$
--- that is especially interesting since the two constituent atoms
have similar charges, which allows one to specify a Bragg reflection
for which the structure factors of Sb and In have equal magnitude but
opposite sign, like in the SMS case. This significantly lowers PXR and
provides a handle to achieve essentially background-free PMR.

Figure~\ref{fig:2} shows our main results, i.e. the emission spectra
as a function of the dimensionless frequency $x$, measured in
$\Gamma/2$. Qualitatively, as expected from Eqs.~(\ref{eq:1}),
(\ref{eq:2}), we find that the peak of the emission occurs at
frequencies where the Cherenkov radiation condition is exactly
fulfilled, i.e. $q' = 0$ and the maximum of the amplitude of the
diffracted wave is reached ($|\alpha_{\mathrm{B}} + \chi_{0}'|$ is
minimal). The asymmetry of the distribution is caused by the fact that
the contribution of the nuclear polarizability to $\chi_{0}'$ changes
its sign when $\omega$ crosses the nuclear resonance frequency
$\omega_0$.

Quantitatively, for electron bunches narrow in the transversal
$z$-direction ($\Delta a = 6.8\cdot10^{-6}\ \mathrm{cm}$ for Spring-8
and $\Delta a = 2.5\cdot10^{-5}\ \mathrm{cm}$ for ESRF), our analysis
predicts that the number of photons that are emitted in the spectral
interval $\Gamma$ ($\Delta x = 2$ in Fig.~\ref{fig:2}) near the
maximum of the distribution is
$N^{\mathrm{Fe}}_{\mathrm{Spring-8}} = 26 157\ \mathrm{cps}$ and
$N^{\mathrm{Fe}}_{\mathrm{ESRF}} = 36 978\ \mathrm{cps}$. For the CsF
crystal, the corresponding number of photons is lower,
$N^{\mathrm{CsF}}_{\mathrm{Spring-8}} = 3074$~cps. The reason is that
the value of the $\xi$ parameter is smaller in this case. Finally, in
the case of the InSb crystal, one obtains
$N^{\mathrm{InSb}}_{\mathrm{Spring-8}} = 3671$ cps. As expected, we
find that the electronic component is strongly suppressed due to the
choice of $(222)$ reflection, when the structure factors of In and Sb
are of an opposite sign. As a result, the PMR paves the way for an
essentially background-free direct spectroscopy of Sb in the energy
domain.

To illustrate its capabilities, in Fig.~\ref{fig:3} we simulate the
spectroscopy of $\alpha$-iron with our source. We find that a
well-resolved spectrum with good contrast can be achieved.

In summary, we have suggested a versatile X-ray source for M\"ossbauer
spectroscopy, based on Parametric M\"ossbauer radiation (PMR) emitted
by relativistic electrons passing through a crystal.  It complements
currently existing M\"{o}ssbauer radiation sources due to its
different qualitative properties. First, the possibility to obtain
collimated photon beams without the need of X-ray optics and
preliminary monochromatization of the radiation. Second, this type of
source is universal and can be realized for a large variety of
M\"{o}ssbauer crystals, including those with forbidden Bragg reflexes,
thus leading to almost background free M\"{o}ssbauer radiation. It
therefore provides a route towards the exploration of M\"ossbauer
spectroscopy beyond the standard isotopes. An interesting perspective
is parasitic operation using dumped electron beams, since PMR converts
the relativistic electrons into resonant X-ray radiation in a cm-scale
crystal.

\begin{acknowledgments}
  \textbf{Acknowledgments}\\
  ODS is grateful to K. P. Heeg, A. Angioi,
  B. Nickerson, S. Kobzak, S. Bragin and D. Bakucz Can\'{a}rio for
  useful discussions.
\end{acknowledgments}

\begin{acknowledgments}
  \textbf{Author contributions}\\
  ODS and IDF initially conceived the project and performed
  calculations. ODS generated the figures. JE proposed the averaging
  procedure over the electron parameters, which vary over the crystal
  length and contributed to the discussion about angular spread of the
  X-rays. ODS, IDF and JE wrote the manuscript. CHK supervised the
  project. All authors contributed to the preparation of the
  manuscript.
\end{acknowledgments}

\section{Appendix A. Details of the calculations}
\label{sec:methods}

\textbf{Solution of Maxwell's equations.} The differential number of
photons $\partial N_{\omega s}/(\partial\omega \partial\Omega)$
emitted in the frequency interval $(\omega$, $\omega+d\omega)$ and in
the solid angle $d\Omega$ is computed in the following
way\cite{skoromnik_parametric_2019, PXR_Book_Feranchuk,
  BARYSHEVSKY1986306}. One starts from inhomogeneous Maxwell's
equations for the Fourier component of the fields, which contain the
current generated by a charged particle. In the case of PXR, the
charged particle moves uniformly, i.e.,
$\vec r(t) = \vec r_{0} + \vec v t$, where $\vec r_{0}$ is the initial
position at $t = 0$. The displacement field $\vec D(\vec r, \omega)$
is related to the electric field $\vec E(\vec r,\omega)$ through the
permittivity tensor
$\epsilon_{\alpha\beta}(\vec r, \vec r_{1},\omega)$, which is defined
in the whole space, but has different expressions inside the crystal
and outside, in vacuum. To facilitate the calculation, we expand the
permittivity inside the crystal in a series over the reciprocal
lattice vectors $\vec g$.

After this, Green's function for Maxwell's equations is defined and
expressed through the solution $\vec E_{\vec k's}^{(-)}$ of
homogeneous Maxwell's equations. Then, Green's function is used to
determine the field generated by the current. This field is then used
in the standard expression for the energy density yielding
\begin{align}
  \frac{\partial^2 N_{\vec n,\omega s}}{\partial \omega \partial
  \Omega} = \frac{e_0^2 \omega }{4 \pi^2 \hbar c^3} \left|\int \vec
  E_{\vec k' s}^{(-)\ast} (\vec r (t),\omega) \cdot \vec v (t)
  e^{\ri\omega t}
  dt\right|^2,\label{eq:5}
\end{align}
where $\vec k' = k\vec r/r$. It is important to note that the solution
$\vec E_{\vec ks}^{(-)}$ of homogeneous Maxwell's equations possesses
an asymptotic behaviour for large $|\vec r|$ as a plane wave and an
ingoing spherical wave. In contrast, when an external electromagnetic
field $\vec E_{\vec ks}^{(+)}$ is diffracted or scattered on a
crystal, it has an asymptotic behavior of a plane wave and an outgoing
spherical wave. However, these two field configurations are related to
each other by the reciprocity theorem~\cite{born2013principles}
$\vec E_{\vec ks}^{(-)} = \vec E_{-\vec ks}^{(+)}$. Thus the actual
problem is reduced to the solution of the diffraction problem to find
the field $\vec E_{\vec ks}^{(+)}$, the usage of the reciprocity
theorem, and the subsequent application of Eq.~(\ref{eq:5}). For this
reason, the actual vector of the emitted photon $\vec k'$ is related
to the vector $\vec k$ of the diffraction problem via
$\vec k' = -\vec k$.

\textbf{Solution of the diffraction problem.} The diffraction problem
is solved within the two-wave approximation of the dynamical
diffraction theory~\cite{authier2001dynamical}, which is valid if two
strong electromagnetic waves are excited in the crystal. The
amplitudes of these waves satisfy a set of homogeneous algebraic
equations
\begin{align}
  \begin{aligned}
    \left(\frac{k^2}{k_0 ^2} -1 - \chi_0 \right) E_{\vec k s}  - c_s
    \chi_{-\vec g} E_{\vec k_g s} &= 0,
    \\
    \left(\frac{k_g^2}{k_0^2} -1 - \chi_0 \right) E_{\vec k_g s}  - c_s
    \chi_{\vec g} E_{\vec k s} &= 0,
  \end{aligned}\label{eq:6}
\end{align}
where $k_{0} = \omega / c$, the incident wave
$\vec E_{\vec ks}^{(+)} = \vec e_{s}E_{\vec ks}$ and the diffracted
wave $\vec E_{\vec k_{g}s}^{(+)} = \vec e_{1s}E_{\vec k_{g}s}$. A
non-trivial solution of this linear homogeneous equation system
exists, if the corresponding determinant is vanishing. This condition
determines the dispersion relation, and its solutions
$\varepsilon_{1s}$ and $\varepsilon_{2s}$ fix the wave vectors
$\vec k_{1,2s} = k_{0}\vec n - k_{0} \varepsilon_{1,2s} \vec N$ of the
diffracted waves. Here, $\vec n$ is the unit vector in the direction
of the incident wave in vacuum. Having found the solutions of the
dispersion equation, one writes down Maxwell's equations in the
crystal and in vacuum and exploits the continuity of the fields at the
crystal-vacuum interface. This fixes the amplitudes of all waves. In
particular, the electromagnetic field responsible for the formation of
PMR equals to
$\vec E_{\vec ks}^{(+)} = \vec e_{1s} E_{g1s} e^{\ri\thp{k_{g}}{r} -
  \ri k_{0}z \varepsilon_{1s}}$, with
$E_{g1s} = c_{s}\chi_{\vec g} / (\alpha_{\mathrm{B}} + \chi_{0})$.

\textbf{Differential number of photons emitted by an electron}. The
integration over the particle trajectory in Eq.~(\ref{eq:5}) with the
law of motion, together with the expression $\vec E_{\vec ks}^{(+)}$
for the electromagnetic field yields
\begin{align}
  \frac{\partial^2 N_{\vec n,\omega s}}{\partial \omega  \partial
  \Omega}
  &= \frac{e_0^2  \omega }{4 \pi^2 \hbar c^5} \sum_{s = \sigma,\pi}
    (\vec e_{1s} \cdot \vec v)^2 \nonumber
  \\
  &\times \left|E_{g1s} L_{g} (1 - e^{- \ri L/L_g})\right|^{2}
    e^{-2 k_{0} |\varepsilon''_{1s} z_{0}|}, \label{eq:7}
\end{align}
where $L_{g} = 1/(k_{0}q)$ is the coherence length and
$q = 1 + (\vec k_{g}\cdot \vec v)/\omega_{0} - \varepsilon_{1s}
v_{z}/c$.

In order to fix a coordinate system and to determine the direction of
the X-ray emission, we for the moment consider an ideal case, in which
the electron velocity does not have any component in the transverse
direction and the minimum of Bragg's condition Eq.~(\ref{eq:2}) is
reached, i.e., $\alpha_{\mathrm{B}} = \gamma^{-2}$.  We align the
$x$-axis parallel to the electron velocity, and the $z$-axis along the
normal $\vec N$ to the crystal surface. In this geometry, the incident
electron beam, as well as the diffracted wave with vector
$\vec k_{g} = \vec k + \vec g$, both propagate along the crystal
surface~\cite{skoromnik_parametric_2019}, such that
$\vec k_{g} \cdot \vec N = 0$.  For a given Bragg reflex, we denote
the projection of the corresponding reciprocal lattice vector $\vec g$
onto the $x-y$ plane by $\vec g_{\bot}$.

The remaining task is to determine the deviations from the Cherenkov
radiation condition $q' = \re q = 0$ and the deviation
$\alpha_{\mathrm{B}}$ from the Bragg's diffraction condition for
non-ideal particle velocities.
For this, we consider electrons with velocities deviating from the
ideal velocity $\vec v_{0} = v \vec e_{x}$. We parameterize these
deviations via
$\vec v = v(\cos\theta_{\mathrm{e}} \vec e_{x} + \vec
\theta_{\mathrm{e}})$, with
$\vec \theta_{\mathrm{e}} = (0, \theta_{\mathrm{e}y},
\theta_{\mathrm{e}z})$ and
$\theta_{\mathrm{e}}^{2} = \theta_{\mathrm{e}y}^{2} +
\theta_{\mathrm{e}z}^{2}$. Analogously, the wave vector of the emitted
radiation
$\vec k = k_{0}(\sin\theta \cos\phi, \sin\theta \sin\phi, \cos\theta)$
will acquire deviations from its ideal direction $\vec k_{0}$.
In order to determine these deviations, we expand the angular
dependence in a Taylor series around the ideal direction $\theta_{0}$
and $\phi_{0}$ up to second order, i.e.,
$\vec k = \vec k_{0} + \vec u_{1} + \vec u_{2}$;
$\vec k^{2} = k_{0}^{2}$; $\thp{k_{0}}{u_{1}} = 0$. As a result of the
deviations from the ideal directions, the quantities $q'$ and
$\alpha_{\mathrm{B}}$ will exhibit corresponding variations
\begin{align}
  q'
  &= (\theta - \theta_{0}) \cos\theta_{0} \cos\phi_{0} - (\phi - \phi_{0})
    \sin\theta_{0}\sin\phi_{0}, \label{eq:8}
  \\
  \alpha_{\mathrm{B}}
  &= -\Big[\gamma^{-2} + (\theta_{\mathrm{e}z} -
    (\theta - \theta_{0}) \sin\theta_{0})^{2} \nonumber
  \\
  &\mspace{40mu}+ (\theta_{\mathrm{e}y} + (\phi - \phi_{0}) \sin\theta_{0}
    \cos\phi_{0} \nonumber
  \\
  &\mspace{100mu}+ (\theta - \theta_{0})
    \cos\theta_{0}\sin\phi_{0})^{2}\Big]. \label{eq:9}
\end{align}

The integration over the emission angles with respect to $\phi$ is
performed in the following manner. First, we apply a variable change
$\phi - \phi_{0} \to q'$. Second, we exploit the fact that the
distribution function is sharply peaked near $\phi = \phi_{0}$, which
allows us to extend the integration range from
$[-\phi_{0}, 2\pi - \phi_{0}]$ to the interval $(-\infty,
\infty)$. Third, since the imaginary part of $q$ is much smaller then
its real part, we can simplify its evaluation by using the value
$\phi$ for the maximum of the intensity.  This intensity maximum is
located at $q' = 0$, which fixes the relation between $\theta$ and
$\phi$.  Thus, we substitute
$\phi - \phi_{0} = (\theta - \theta_{0})\cot\theta_{0} \cot\phi_{0}$
in the imaginary part of $q$. Finally, we perform the integration with
the help of the residue theorem yielding Eq.~(\ref{eq:3}).

In addition, it is important to note that the electron velocity spread
in the transversal $y$-direction, which is
typically~\cite{Spring8Parameters, ESRFParameters} much larger then
the corresponding spread in the $z$-direction, does not influence the
emitted number of photons. This is due to the independence of the
photon distribution function of the initial position $y_{0}$ of the
electron for the case of the EAD geometry.
\begin{figure}[t]
  \includegraphics[width=0.48\textwidth]{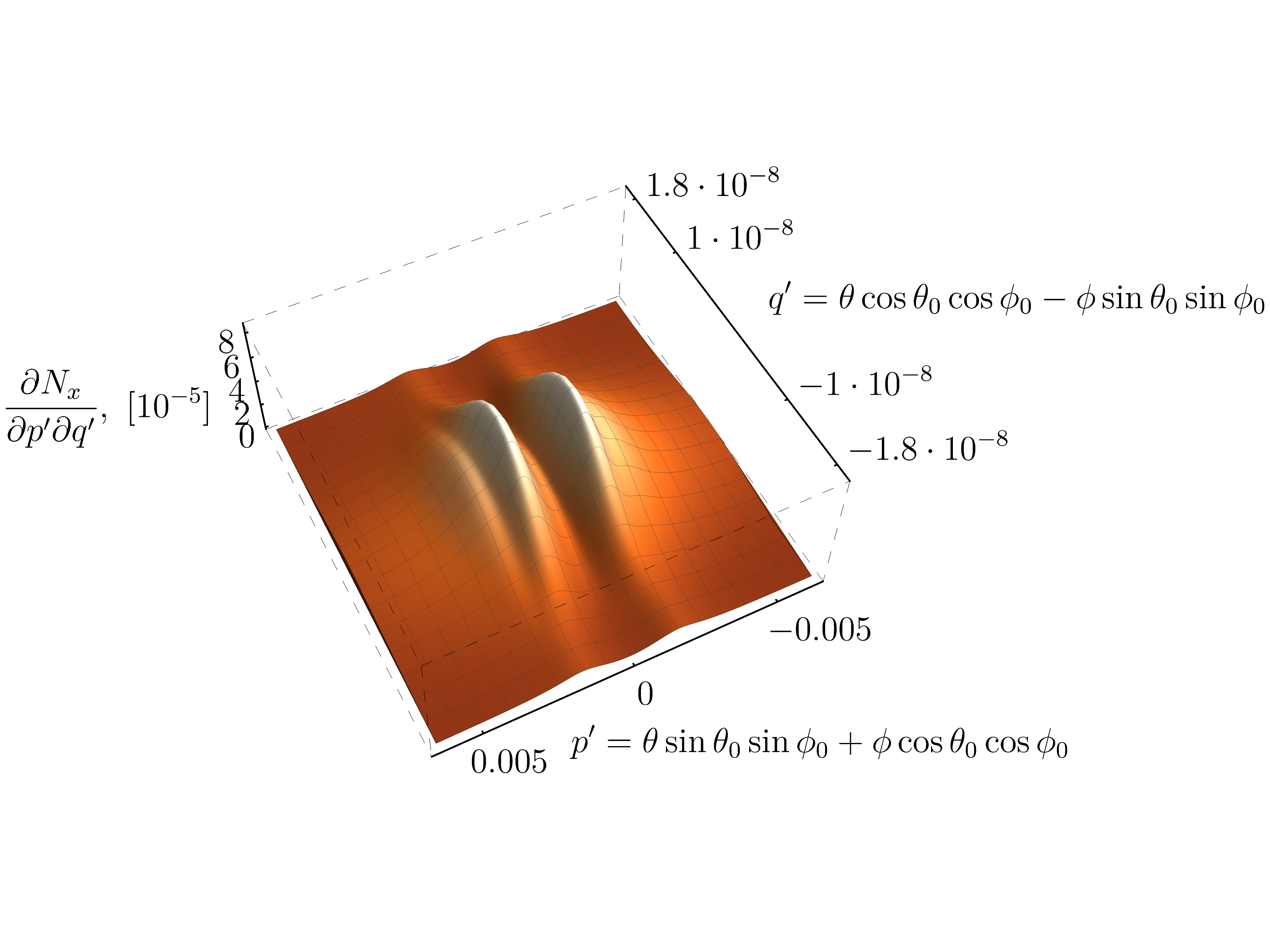}
  \caption{\label{fig:4}{\bf Angular distribution of the PMR
      intensity.} The figure shows results for the Spring-8 facility
    as a function of
    $p' = \theta \sin\theta_{0}\sin\phi_{0} + \phi
    \cos\theta_{0}\cos\phi_{0}$ and
    $q' = \theta \cos\theta_{0}\cos\phi_{0} - \phi
    \sin\theta_{0}\sin\phi_{0}$ parameters for $(011)$ reflection from
    $\alpha$-iron crystal. When $q' = 0$, the Cherenkov radiation
    condition is exactly fulfilled. The angular distribution is
    averaged over the electron bunch characteristics with the
    parameters: $\Delta a = 6.8\cdot 10^{-6}\ \mathrm{cm}$,
    $\Delta\theta_{\mathrm{e}z} = 10^{-4}\ \mathrm{rad}$,
    $\Delta\theta_{\mathrm{e}y} = 10^{-3}\ \mathrm{rad}$ and
    normalized to one electron and a measurement time of one second.}
\end{figure}

\textbf{Averaging over the electron bunch parameters and multiple
  electron scattering.} The velocity distribution of experimentally
available electron bunches is characterized via the emittance
$\epsilon_{y,z} = \Delta a \Delta \theta_{\mathrm{e}y,z}$, with
$\Delta a$ the transversal and $\Delta\theta_{\mathrm{e}y,z}$ the
angular spreads. As a result, we need to average the spectral-angular
emission distribution obtained for a single electron over the
parameters of the entire electron bunch. This is achieved by
convolving the emission distribution with the electron angular
distribution $\theta_{\mathrm{e}y}$, $\theta_{\mathrm{e}z}$ and the
initial $z_{0}$ coordinate distribution. We consider Gaussian
distribution functions for the electron beam parameters given by
\begin{align}
  G
  &(\theta_{\mathrm{e}z}, \theta_{\mathrm{e}y}, z_0) = C
    F(\theta_{\mathrm{e}z}, \theta_{\mathrm{e}y}, z_0), \label{eq:10}
  \\
  F &=  \exp[-((\theta_{\mathrm{e}z}-\theta_{0\mathrm{e}})^{2} /
      (\theta_{\mathrm{s}}^{2} + \Delta\theta_{\mathrm{e}z}^{2}) +
      \theta_{\mathrm{e}y}^{2} / (\theta_{\mathrm{s}}^{2} +
      \Delta\theta_{\mathrm{e}y}^{2}))]
      \nonumber
  \\
  &\mspace{40mu}\times \exp[- ( z_{0}  - a_{0})^{2} / \Delta a^{2}].
    \nonumber
\end{align}
The constant $C$ is a normalization constant, chosen such that the
total intensity corresponds to the single electron case, i.e.,
$\int d \theta_{\mathrm{e}z} d \theta_{\mathrm{e}y} d z_0
G(\theta_{\mathrm{e}z}, \theta_{\mathrm{e}y}, z_0) = 1$. The angle
$\theta_{0\mathrm{e}}$ is the mean incidence angle of the electron
bunch on the crystal. The angle
$\theta_{\mathrm{s}}^{2} = (E_{\mathrm{c}}/E)^{2}(L/L_{\mathrm{R}})$
characterizes multiple electron scattering\cite{ter1972high}, with
$E_{\mathrm{c}} \approx 21\ \mathrm{MeV}$, $L$ the crystal length and
$L_{\mathrm{R}}$ the radiation length. For Fe, the latter is
$L^{\mathrm{Fe}}_{\mathrm{R}} = 1.757\ \mathrm{cm}$
\cite{PhysRevD.98.030001}, for CsF
$L^{\mathrm{CsF}}_{\mathrm{R}} = 2.227\
\mathrm{cm}$\cite{PhysRevD.98.030001} and for InSb
$L^{\mathrm{InSb}}_{\mathrm{R}} = 3.701\
\mathrm{cm}$\cite{PhysRevD.98.030001}.

In addition, we perform the averaging over the beam transversal
spread. For this we consider that the beam divergence is not constant
along the crystal length, but is given by
\begin{align}
  \sigma(s) = \sqrt{\Delta a^{2} + \epsilon^{2} \frac{s^{2}}{\Delta
  a^{2}}}, \quad -L/2 \leq s \leq L/2 \label{eq:13}
\end{align}
instead. Here $\Delta a$ is the transversal spread in the focus
center. Therefore, in the actual calculation we vary $\Delta a$ in
Eq.~(\ref{eq:10}) from $\sigma(-L/2)$ to $\sigma(L/2)$ in $R-1$ steps
and average the resulting spectra over the resulting values, i.e.
$\partial N_{x}/\partial x = 1/R\sum_{i=0}^{R-1} \partial
N^{i}_{x}/\partial x$. Therefore, in all figures above the given value
$\Delta a$ corresponds to the value in the center of the focus of the
beam.

\textbf{Temperature effects.} The effects of lattice vibrations on the
crystal polarizabilities\cite{afanasev_suppression_1965} due to the
temperature are taken into account with the help of the Debye–Waller
factor $e^{-W(\vec k, \vec k_{g})}$, which for temperatures much
smaller than the Debye temperature $\Theta_{\mathrm{D}}$ and an
isotropic cubic crystal is expressed as
\begin{align}
  W = \frac{3\hbar^{2} g^{2}}{8m
  k_{\mathrm{B}}\Theta_{\mathrm{D}}}
  \left[1 + \frac{2\pi^{2}}{3}
  \left(\frac{T}{\Theta_{\mathrm{D}}}\right)^{2}\right], \label{eq:11}
\end{align}
where $k_{\mathrm{B}}$ is the Boltzmann constant, and $m$ the mass of
the resonant isotope. For $\alpha$-iron the Debye temperature
$\Theta^{\mathrm{Fe}}_{\mathrm{D}} = 470^{\circ}\ \mathrm{K}$, for CsF
$\Theta^{\mathrm{CsF}}_{\mathrm{D}} = 109^{\circ}\ \mathrm{K}$ and for
InSb $\Theta^{\mathrm{InSb}}_{\mathrm{D}} = 163^{\circ}\ \mathrm{K}$.

\textbf{Numerical values of the parameters used in the calculations.}
We choose the most intense reflection for the $\alpha$-iron crystal,
namely, the $(011)$ reflection. For this reflection we employ the
following parameters, taken from the X-ray database
\cite{StepanovXrayWebServer}
\begin{equation}
  \begin{aligned}
    &\hbar\omega_{\mathrm{B}} = 14.41\,\mathrm{keV},& &k_{0} =
    7.35\times10^{8}\,\mathrm{cm}^{-1},
    \\
    &\chi_{0\mathrm{e}}' = -0.15\times10^{-4},& &\chi_{0\mathrm{e}}''
    = 0.69\times 10^{-6},
    \\
    &\chi_{\vec g\mathrm{e}}' = -0.10\times 10^{-4},& &\chi_{\vec
      g\mathrm{e}}'' = 0.67\times 10^{-6}.
  \end{aligned}\label{eq:12Fe}
\end{equation}

The $\alpha$-iron crystal has cubic crystalline structure with
inter-planes distance $d = 2.87\times 10^{-8}\ \mathrm{cm}$. In
addition, we assume it to be enriched to $90\%$ with the resonant
M\"{o}ssbauer isotope $_{26}^{57}\mathrm{Fe}$, which has the natural
decay width $\Gamma = 4.66\times 10^{-12}\ \mathrm{keV}$. The
coefficient of internal conversion $\alpha_{\mathrm{C}} = 8.56$ and
the structure factor $S(\vec g) = 2$ for $_{26}^{57}\mathrm{Fe}$.

For CsF, we employ the $(111)$ reflection with the
parameters~\cite{StepanovXrayWebServer}
\begin{equation}
  \begin{aligned}
    &\hbar\omega_{\mathrm{B}} = 80.997\,\mathrm{keV},& &k_{0} =
    4.10\times10^{9}\,\mathrm{cm}^{-1},
    \\
    &\chi_{0\mathrm{e}}' = -0.25\times10^{-6},& &\chi_{0\mathrm{e}}''
    = 0.42\times 10^{-8},
    \\
    &\chi_{\vec g\mathrm{e}}' = -0.15\times 10^{-6},& &\chi_{\vec
      g\mathrm{e}}'' = 0.39\times 10^{-8}.
  \end{aligned}\label{eq:12CsF}
\end{equation}
The CsF crystal has a cubic crystalline structure with inter-planar
distance $d = 6.008\times 10^{-8}\ \mathrm{cm}$. The natural decay
width of the $_{55}^{133}\mathrm{Cs}$ isotope is
$\Gamma = 72.77\times 10^{-12}\ \mathrm{keV}$, the internal conversion
coefficient $\alpha_{\mathrm{C}} = 1.72$, and the structure factor
$S(\vec g) = 4$.

For InSb, we employ the $(222)$ reflection with the
parameters~\cite{StepanovXrayWebServer}
\begin{equation}
  \begin{aligned}
    &\hbar\omega_{\mathrm{B}} = 37.133\,\mathrm{keV},& &k_{0} =
    1.88\times10^{9}\,\mathrm{cm}^{-1},
    \\
    &\chi_{0\mathrm{e}}' = -0.15\times10^{-5},& &\chi_{0\mathrm{e}}''
    = 0.72\times 10^{-7},
    \\
    &\chi_{\vec g\mathrm{e}}' = 0.17\times 10^{-7},& &\chi_{\vec
      g\mathrm{e}}'' = -0.46\times 10^{-8}.
  \end{aligned}\label{eq:12InSb}
\end{equation}
The CsF crystal has cubic crystalline structure with inter-planar
distance $d = 6.4789\times 10^{-8}\ \mathrm{cm}$.  The isotope
$_{51}^{121}\mathrm{Sb}$ has a natural decay width
$\Gamma = 0.13\times 10^{-6}\ \mathrm{eV}$, the internal conversion
coefficient $\alpha_{\mathrm{C}} = 11.11$. The structure factors are
$S(\vec g) = -4$ for $_{51}^{121}\mathrm{Sb}$ and $S(\vec g) = 4$ for
$_{49}\mathrm{In}$.

Regarding the electron bunch parameters, we have investigated two
accelerator facilities, namely Spring-8 \cite{Spring8Parameters} with
electron beam energy $8000\,\mathrm{MeV}$, and
ESRF\cite{ESRFParameters} with electron beam energy
$6030\ \mathrm{MeV}$. The Spring-8 facility provides electron beams
with natural vertical emittance
$\epsilon = 6.8\times 10^{-10}\,\mathrm{cm}\times \mathrm{rad}$, while
the ESRF facility has a vertical emittance
$\epsilon = 2.5\times 10^{-9}\,\mathrm{cm}\times\mathrm{rad}$. For all
simulations, the angular spread in the horizontal $y$-direction
$\Delta\theta_{\mathrm{e}y}$ was taken to be $10^{-3}\ \mathrm{rad}$.

\textbf{Direction and divergence of the X-ray emission.}
Table~\ref{tab:1} summarizes the angles characterizing the vector
$\vec k$ which determines the X-ray emission direction. The actual
emission is happening in the direction $\vec k' = -\vec k$. Possible
values for the angles $\phi_{0}$, $\psi_{0}$ range from $-\pi$ to
$\pi$, and values for the angle $\theta_{0}$ range from $0$ to $\pi$.
\begin{table}[t]
  \centering
  \begin{tabular}{|c|c|c|c|}
    \hline
    Crystal & $\theta_{0}$ & $\phi_{0}$ & $\psi_{0}$
    \\
    \hline
    $\alpha$-iron & 107.468 & -162.532 & 107.468
    \\
    \hline
    CsF & 91.46 & -177.94 & 91.55
    \\
    \hline
    InSb & 95.91 & -171.62 & 96.28
    \\
    \hline
  \end{tabular}
  \caption{The angles $\theta_{0}$, $\phi_{0}$ of a spherical
    coordinate system together with the angle $\psi_{0}$, which
    determine the direction of emission and the orientation of the
    crystal with respect to the particle velocity (see
    Fig.~\ref{fig:1}).}\label{tab:1}
\end{table}
In the angular distribution of the emitted radiation shown in
Fig.~\ref{fig:4}, two qualitatively different scales can be
observed. A narrower first scale arises from the Cherenkov radiation
condition.  It is satisfied exactly at $q' = 0$. In this case, the
angular width is defined through the coherent length $L_{\mathrm{g}}$
and the width of $q' \sim (k_{0}L_{\mathrm{g}})^{-1} \sim
10^{-8}$. The second direction, which is perpendicular to $q'$ is
characterized via a variable $p'$. This variable is associated to the
maximum of the diffracted wave, and is of the order
$p' \sim \sqrt{|\chi_{0}''|} \sim 10^{-3}$. Consequently, the PMR is
concentrated around the direction given by the vector
$\vec k'_{0} = -\vec k_{0} = (g_{x} + \omega_{0}/v, g_{y}, g_{z})$.
% within the solid angle
% $\Delta\Omega \approx 10^{-3}\times 10^{-8}\ \mathrm{rad}^{2}$.

Due to the finite crystal size $\sim 1\ \mathrm{mm}$, the beam of PMR
seen by the detector has at least a width given by the crystal size,
projected onto the plane defined by the vector normal to the
detector. The beam divergence is of order
$\Delta\Omega \approx 10^{-3}\times 10^{-8}\ \mathrm{rad}^{2}$ is
defined via the angular divergence of the emitted gamma quanta
($\simeq \gamma^{-1}$). Therefore, the target and detector should
ideally be comparable or larger than the crystal size.

\textbf{Calculation of the absorption spectrum.} The absorption
spectrum is computed as
\begin{align}
  N(\omega_{s})
  &= N_{\mathrm{e}} \int \left(\frac{I(\omega) +
    I(-\omega)}{2} - I_{\mathrm{B}}\right) \nonumber
  \\
  &\mspace{60mu}\times e^{-k_{0}D |\im \chi_{0}(\omega - \omega_{s})|}
    d\omega, \label{eq:14}
\end{align}
where $I_{\mathrm{B}}$ is the electronic part of the intensity and
$\chi_{0}(\omega - \omega_{s})$ includes both the electronic and the
nuclear polarizabilities. Here $N_{\mathrm{e}}$ is the number of
electrons per second.

\bibliography{pxr_bibliography}

\end{document}